AR/5

# Phenomenological Theory of Superfluidity and Superconductivity

Mario Rabinowitz
Armor Research; lrainbow@stanford.edu
715 Lakemead Way, Redwood City, CA 94062-3922

## Abstract

Quantum condensation is used here as the basis for a phenomenological theory of superfluidity and superconductivity. It leads to remarkably good calculations of the transition temperatures $T_c$ of superfluid $^3He$ and $^4He$, as well as a large number of cuprate, heavy fermion, organic, dichalcogenide, and bismuth oxide super- conductors. Although this approach may apply least to the long coherence length metallics, reasonably good estimates are made for them, and chevral superconductors. $T_c$ for atomic H is estimated. $T_c$ can be calculated as a function of number density or density of states, and effective mass of normal carriers; or alternatively with the Fermi energy as the only input parameter. Predictions are made for a total of 26 superconductors and 4 superfluids. An estimate is also made for coherence lengths.

## 1. INTRODUCTION

The enigma of high temperature superconductivity has produced a deluge of conceivable theories too numerous and diverse to list here. This plethora of theories are all plagued by the same problem that no critical temperatures ($T_c$) can be calculated until both the interaction and its strength are known. In the case of the low $T_c$ metallics where the phonon interaction and the BCS theory (Bardeen *et al* , 1957) are well formulated, calculations are exceedingly difficult (Carbotte 1990) and consume considerable time and energy due to their complexity. In conventional BCS superconductors, Cooper pairs bound by phonon interaction form with zero total spin and zero total angular momentum (s-wave), and BCS $T_c$ calculations give satisfactory agreement with experiment for monatomic metals and for some ordered alloys. However, they are inapplicable to materials beyond the typical parameter ranges.

BCS may also be inapplicable to a large number of both proposed and known unorthodox superconductors and superfluids. It appears inappropriate to the quasi-one-dimensional organic superconductors with side chains conceived by Little (1964), and possibly the hydrides proposed by Overhauser (1987). The recently discovered unorthodox superconductors such as the cuprates, bismuth oxides, organics, and at least some heavy fermion superconductors fall outside the domain of the BCS theory because they have non-phonon pairing interactions. Recently, Hasuo et al (1993) report evidence of a Bose-Einstein (B-E) condensation of biexcitons in CuCl. The presence of a weak population of biexcitons in the K= 0 ground state seems to play an important role in the nucleation of the superfluid transition.

In superfluid $^3$He the non-phonon spin-flip mediated pairing has been clearly identified, and pairing is in the triplet state of spin 1 with p-wave total orbital angular momentum (Leggett 1975). There is evidence for d-wave rather than s-wave pairing in the cuprates (Coffey and Coffey 1993). Pairing in some heavy fermion materials appears to be primarily d-wave mediated by antiferromagnetic fluctuations with the triplet spin state as for $UPt_3$. There was a time when anything other than zero spin and zero angular momentum would have been inconceivable to many.

With rare exceptions such as the prediction of superconductivity in high pressure phases of materials such as Si by Cohen and his colleagues (Erskine *et al* , 1986), the discovery of unorthodox as well as conventional superconductors has come neither from basic theory, nor from guidance from the microscopic principles of superconductivity. One generalization that applies to the new materials is that they are increasingly complex. This may just be an example of the power of large numbers since the number of possibilities increases greatly with complexity as the building blocks go from unaries to binaries to ternaries to quaternaries, etc. However, there may be an optimum degree of complexity. London (1937) speculated that life may depend on high $T_c$ (Rabinowitz 1990a). Superconductivity in the simplest materials may not at all be representative of superconductivity in its entirety. The only significance that the earliest discoveries of superconductivity were in the simplest materials may simply be because they were the easiest to work with, rather than that this form of superconductivity is the most prevalent. A possibility worth considering, though it does not appear to be verifiable, is that all conducting materials could become superconducting if taken to sufficiently low temperatures.

A seminal generalization is that to be superconductors or superfluids regardless of their simplicity or complexity and the nature of the pairing mechanism, they must all exhibit quantum condensation in phase space. In my opinion, this is the condition that the entire diversity of materials and theories must obey. My position is that this condition is sufficient for **approximate** calculations of $T_c$, without the need for knowing the interaction. The interaction-free approach initiated by Rabinowitz (1987, 1988, 1989a,b, 1990b, 1993a,b) leads to simple formulas which work exceedingly well with great computational ease. They can be of benefit as a guide to the development of more rigorous and complete theories, as well as in the quest for new and better superconductors.

## 2. QUANTUM CONDENSATION

Quantum condensation in phase space leads to the formation of a quantum fluid which can exhibit macroscopic quantum effects such as superconductivity and superfluidity. In simple terms, it is the multiple occupancy (degeneracy) of available energy levels. Previously, I used the condition (Rabinowitz, 1989a) that the thermal de Broglie wavelength $\lambda$ must be greater than the inter-particle Boson spacing d. For greater generality, and to be more explicit, let us allow that there may be more than one kind of Boson associated with different spin states

$$\lambda \geq 2d = 2\left(\frac{n_B}{g}\right)^{-1/3} \tag{1}$$

where $n_B$ is the total 3D number density of bosons, and g is the spin degeneracy so that $n_B / g$ is the number density of bosons of a given kind.

Let us see that equation (1) is equivalent to multiple occupancy of energy levels. Briefly, without going into a lot of extraneous detail, for a free carrier model the 3D density of states for both spins (per unit energy per unit volume) is

$$G_{3D} = \frac{E^{1/2}}{2\pi^2}\left(\frac{2m}{h^2}\right)^{3/2} , \tag{2}$$

where m is the effective mass, and E is a given energy level. We require that the number of specific particles per energy level > 1:

$$\frac{V(n_B / g)}{V\int G_{3D}dE} = \frac{n_B}{g}\left[\frac{3\pi^2}{E^{3/2}}\left(\frac{h^2}{2m}\right)^{3/2}\right] = \frac{3\lambda^3}{8\pi}\left(\frac{n_B}{g}\right) > 1. \tag{3}$$

Equation (3) implies

$$\lambda \gtrsim 2\left(\frac{n_B}{g}\right)^{-1/3} = 2d \tag{4}$$

in agreement with equation (1). Alder and Peters (1989) using quantum Monte Carlo calculations have shown the variability of the coefficient linking $\lambda$ and d.

## 3. TRANSITION TEMPERATURES

### 3.1. Superfluid Liquid $^4$He

The de Broglie wavelength at the thermal energy $(f/2)\kappa T_c$ is

$$\lambda = h / \left[2m_B (f/2)\kappa T_c\right] . \qquad (5)$$

$m_B$ is the effective boson mass, f is the degrees of freedom, and $\kappa$ is the Boltzmann constant. Combining equations (1) and (5) we obtain

$$T_c^{BE} \le \frac{h^2}{12m\kappa}\left(\frac{n_B}{g}\right)^{2/3} \qquad (6)$$

in close agreement with the B-E condensation temperature (Betts, 1969) where the 12 is replaced with 11.92. Equation (6) works well for $^4$He which is a boson of zero spin with $g = 1$, $m \approx 6.7 \times 10^{-24}$ gm , and $n_B = 2.2 \times 10^{22}/cm^3$ , yielding $T_c^{BE} \approx 3$ K in good agreement with the superfluid transition temperature of 2.17 K for L$^4$He (Rabinowitz, 1993a).

### 3.1. Superfluid Atomic H

Helium is unique among the elements as the only one exhibiting superfluidity since it remains in the liquid state down to the lowest temperatures obtainable because of its very weak interaction, and because its zero point amplitude is high enough to keep it from solidifying. Although hydrogen has a higher zero point energy, it solidifies because it has much stronger interaction.

Since hydrogen is a boson, let us see roughly at what temperature $LH_2$ might become a superfluid if it could remain a fluid. Equation (6) says that the best we could hope for would be $T_c^{BE} \sim 6$ K for $n \sim 2 \times 10^{22}/cm^3$ and $m = 2(1.67 \times 10^{-24}$ gm). Since this is below the solidification temperature for $H_2$ of 14.1 K, superfluidity appears out of the question for $LH_2$. However, recent cooling achievements for H atoms of $T \approx 100$ μK (Doyle *et al* , 1991) at a density of $n = 8 \times 10^{13}/cm^3$ present two exciting possibilities for superfluidity. For the singlet state, at this density $T_c^{BE} \sim 29.5$ μK, not that much lower than what has already been achieved. For a triplet state, $T_c^{BE} \sim 14.2\ \mu K$.

### 3.2. Superfluid Liquid $^3He$

Using this paradigm, we can derive $T_c$ for superfluid $^3He$. This derivation is more general and different than previously done by Rabinowitz (1993a). $^3He$ is a fermion of spin 1/2. It obeys the Pauli exclusion principle, and hence only a fraction of the fermions $\sim \kappa T_c / E_F$ are incipient Cooper pairs at $T_c$, where

$$E_F = \left[\frac{\hbar^2 n^{2/3}}{8m}\right](3/\pi)^{2/3} \tag{7}$$

is the Fermi energy, m is the fermion effective mass, and n is the number density of fermions. This is similar to electrons in which only $\sim \kappa T/E_F$ participate in the electron specific heat. The number of fermions involved is the number of k values within a shell $\Delta k$ of energy width

$$\kappa T_{cF} \approx \frac{\hbar^2 k_F^2}{2m} - \frac{\hbar^2 (k_F - \Delta k)^2}{2m}, \tag{8}$$

where $k = 1/\lambda\!\!\!^- = p/\hbar$ is the wave vector. For a roughly spherical Fermi surface the number density of bosons of a specific kind is

$$n_B = \frac{1}{2}\frac{\Delta k(4\pi k_F^2)}{\frac{4}{3}\pi k_F^3 g} n = \frac{3\Delta k}{2k_F g} n\ . \tag{9}$$

Combining equations (8) and (9),

$$n_B = \frac{3\kappa T_{c3}}{4E_F g} n. \tag{10}$$

For layered materials the Fermi surface is cylindrical, yielding a coefficient of 1/2 instead of 3/4 in eq. (10). However if the fraction of fermions $\sim (3/2)kT_c/E_F$ rather than $\sim kT_c/E_F$, then a cylindrical Fermi surface would also yield eq.(10).

For $^3He$ pairs, $\lambda$ is given by equation (5) with $m_B = 2m$, and $f \geq 3$ in 3D.

Combining equations (1), (5), (7) and (10)

$$T_{c3} = 2.77 \times 10^{-3}\left[\frac{\hbar^2 n^{2/3}}{m\kappa g^2}\right] = 2.28 \times 10^{-2}\left(\frac{T_F^{3D}}{g^2}\right), \tag{11}$$

where $T_F^{3D} = E_F/\kappa$.

For $^3He$ pairs, g = 3, since the spin = 1. One may calculate the number density of $L^3He$ as $n = 2.36 \times 10^{22}/cm^3$ at the solid liquid interface from both the independent data of Wheatley (1975) at 34.36 bar and of Greywall (1986) at 34.39 bar. At these pressures, Wheatley gives $m = 6.22\ m_{^3He}$, and Greywall $m = 5.85\ m_{^3He}$. Using Wheatley's data, equation (11) yields $T_{c3} = 2.6$ mK in excellent agreement with his experimental value of 2.6 mK. Using Greywall's data, equation (11) yields $T_{c3} = 2.8$ mK also in excellent agreement with his experimental value of 2.5 mK. A private communication (1993) by D. D. Osheroff indicates that the originally measured value of $T_{c3} = 2.6$ mK (Osheroff, 1972), should be corrected to $T_{c3} = 2.5$ mK.

Original predictions of $T_c$ were ~0.1 K, and then fell to $10^{-6}$-$10^{-9}$ K, when the $^3He$ superfluid transition was not found down to 0.01 K (Rabinowitz 1993). My theory makes the prediction that the superfluid transition temperature for dilute solutions of $^3He$ in superfluid $^4He$ are ~ 1- 10 μK (Rabinowitz, 1993a) Winterberg (1989) has speculated on the possibility of a superfluid B-E condensation of photons.

## 4. HIGH TEMPERATURE SUPERCONDUCTIVITY

We can use eq. (11) as the starting point for deriving $T_c$ for high-$T_c$ superconductors as well as other superconductors which exhibit a 2D layered structure. Combining equations. (1), (7), (10) and (11) with f = 2 in eq. (5) yields

$$T_{c2} = 4.15 \times 10^{-3}\left[\frac{h^2 n^{2/3}}{m\kappa g^2}\right] = 5.22 \times 10^{-2}\left[\frac{T_F^{2D}}{n^{1/3}\delta g^2}\right] = 3.42 \times 10^{-2}\left[\frac{T_F^{3D}}{g^2}\right], \qquad (12)$$

where g = 1 for singlet pairing (spin = 0) superconductors, $T_F^{2D} = E_F^{2D}/\kappa$, and $\delta$ is the average plane spacing. For direct input of experimental data, eq. (12) can also be written in terms of the three dimensional density of electronic states in the normal phase for both spins $G_{3D}(\mu)$ at the chemical potential $\mu$ (Fermi energy):

$$T_{c2} = 2.71 \times 10^{-5}\left[\frac{h^6 \left[G_{3D}(\mu)\right]^2}{\kappa m^3 g^2}\right]. \qquad (13)$$

However eq. (12) in terms of the single input parameter $T_F^{3D}$ is the simplest possible expression for $T_c$, when it is available. In my theory, the quantity $h^2 n^{2/3}/m\kappa$ always enters in this form, making possible its replacement with $T_F^{3D}$. Unfortunately, a rich source of valuable data compiled by Harshman and Mills (HM) (1992) only lists $T_F^{3D}$ for a small number of superconductors. Instead they list most in terms of $T_F^{2D}$ which is why equation (12) is also given in terms of $T_F^{2D}$.

For the metallic superconductors, we can make a small correction to the equation for $T_c$ previously derived (Rabinowitz 1989b equation (5)) . The 3/4 in equation (10) above enters in as $(3/4)^2$ times the original expression and including g with f = 3 yields

$$T_{cm} = 4.07 \times 10^{-5}\left[\frac{h^2 n^{2/3}}{m\kappa g^2}\right] = 3.26 \times 10^{-4}\left[\frac{T_F^{3D}}{g^2}\right]. \qquad (14)$$

Table I tabulates the predicted transition temperatures $T_c^{pred}$ and the necessary input data for a wide variety of superconductors for comparison with the experimental values $T_c^{exp}$. $^3$He is calculated from equation (11) using Greywall's data. H and $^4$He are determined from equation (6). Sixteen superconductors (No's.6-9 &19-30) are calculated from equation (12) inclusive of all the cuprate, organic, bismuth oxide, dichalcogenide superconductors from data in HM. No.5 (heavy fermion with g=3), and No's. 17 & 18 are calculated from eq.(11) as they are 3D superconductors. The remaining seven metallics are determined using eq. (14). In all cases every single superconductor in HM is presented for which there is sufficient data to make a calculation, which accounts for all the entries except H, He, and the six pure metals. The $E_F$ for the six pure superconducting metals can be
found in most solid state textbooks. The data indicates that for the metallics my simple theory requires modification with increased convolution. Although three significant figures are not usually warranted, they are presented in case they might be needed for computational purposes. The agreement between $T_c^{pred}$ and $T_c^{exp}$ is exceptionally good for 21 out of 28 superconductors-fluids. $T_c^{exp}$ is not known for two entries. The agreement is even better than the table shows, when $T_c$ is calculated directly from n and m -- possibly because of different uncertainties in $T_F^{2D}$ and $\delta$ than ascribed to them.

## 5. COHERENCE LENGTHS

The concept of coherence applies both to particles and to waves. From a classical point of view, we could focus on the coherent motion of electron pairs in superconductivity or neutral carriers in superfluidity. In the BCS theory, the correlated motion of the electrons in a Cooper pair with equal and opposite momenta allow the center of mass to move undisturbed because as one electron scatters the other is required to scatter in the opposite direction. Quantum mechanically the concept of coherence applies to a correlation between phases of the wave function at all points in space. The particles have condensed into
the ground state described by one wave function. It is as if they were all acting like a single body. With the low energy states filled, it is as if there were no available states to fall down to, so scattering cannot lead to energy loss.

In simple physical terms, the concept of coherence length, $\xi$, can be looked at in three different ways. The simplest way is to think of $\xi$ as the rms distance between the electrons in a Cooper pair as they oscillate $180^o$ out of phase about their center of mass. A second physical meaning of $\xi$ relates to the quantum of flux (fluxoid) in a superconductor or vortex in a superfluid, with radius ~ $\xi$.

The third is related to the fact that the superconducting electrons are in a more ordered (more coherent) state than the normal electrons.
Pippard identified the change in density, $n_s$, of superconducting electrons with this order. In the metallics $\xi \sim 10^3 - 10^4$ Å implying that there are billions of electron pairs within a volume $\xi^3$. The long $\xi$ with large overlap of pairs makes pair-pair interactions greatest in the metallics, and least applicable to the interaction-free approach. In the cuprates $\xi \sim$ 5- 20Å, so there are only a small number of electron pairs within a coherence volume with little pair-pair interaction.

Although the primary motivation of my approximate theory is to calculate transition temperatures, a fringe benefit is that it can also yield approximate values of the low temperature coherence length $\xi$ that agree roughly with the metallics, and quite well with the cuprates. In simple terms the uncertainty principle $\delta E\,\delta t \geq \hbar / 2$ can be applied in the region of the low temperature energy gap $2\Delta = \delta E$, and $\delta t = \xi / v_F$ where $v_F$ is the Fermi velocity. This implies that $\xi = \hbar v_F / 4\Delta$. Rigorously one would get,

$$\xi = \frac{\hbar v_F}{\pi \Delta}. \tag{15}$$

For the metallics BCS has $2\Delta = 3.52kT_c$. Combining this with equations (7) and (14) in (15):

$$\xi_{met} = 700 n^{-1/3}. \tag{16}$$

With $n \sim 10^{23}/cm^3$, equation (16) gives $\xi_{met} \sim 2000$ Å in fair agreement with the experimental range of ~$10^3$ - $10^4$ Å .

For the cuprates $2\Delta \sim 8kT_c$ (Rabinowitz 1989a). Combining this with equations (7) and (12) in (15):

$$\xi_{cup} = 3n^{-1/3} \tag{17}$$

With $n \sim 10^{22}/cm^3$, equation (17) yields $\xi_{met} \sim 14$ Å in excellent agreement with the experimental range of ~5 - 20 Å .

## 6. RETROSPECTION AND CONCLUSION

Although Ogg (1946) first proposed electron pairs for superconductivity, electron pairs should rightfully be called Cooper pairs. The pairing idea was proposed many times (Blatt 1964) long before Cooper (1956) proposed it. Nevertheless, Cooper was the first to quantify the problem of two Fermions interacting in the presence of a filled Fermi sea of N - 2 other Fermions.

My analysis clearly differs from that of BCS (1957). B-E condensation approaches to superconductivity were first presented by Schafroth (1955), and Schafroth, Butler, and Blatt (SBB) (1957); and more recently by Friedberg and Lee (1989). These approaches differ from mine, and appear not to have been successful in predicting $T_c$'s. (Standard approaches have also been unsuccessful with high-$T_c$'s.) I was not aware of the SBB work when my earlier papers were written. They showed that if the size of the electron pairs is less than the average distance between them, and if other conditions are fulfilled, the system has properties similar to that of a charged Bose-Einstein gas, including a Meissner effect and a critical temperature of condensation. However, the SBB approach did not lead to $T_c$ predictions, and is much more elaborate than my theory.

The original BCS paper (1957) was not without some shortcomings.It did not predict the mixed state of type II superconductivity. It incorrectly predicted that a small amount of impurity scattering would greatly reduce $T_c$. Nevertheless, it proved to be extremely valuable.

The simplicity of my theory is both a source of strength and weakness. The simplicity invariably leads to limitations. Nevertheless it has impressive power to correlate data over a wide range without explicitly introducing a pairing mechanism. Perhaps this should not be entirely surprising, as $T_c$ itself is a measure of the interaction strength. In this model, $T_c$ enters into the equations in two different ways so that it is possible to solve for $T_c$. It is a kind of self-consistency requirement that the number of pairs that B-E condense (i.e. those having a de Broglie wavelength comparable to their average separation) is proportional to the condensation temperature itself. This may be why the transition temperature can be obtained without prior knowledge of the interaction mechanism, presenting a necessary but not sufficient condition for superconductivity. Even though thermodynamically there is only one $T_c$ for a given material, $T_c \propto 1/m$ suggests that a significantly higher effective mass in the c direction for layered 2D materials than in the ab plane implies that there is an intrincically lower $T_c^c$ in the c direction than $T_c^{ab}$ in the ab planes. This would be like having parallel superconducting sheets with normal connections in the perpendicular direction for $T_c^{ab} > T > T_c^c$. However, coupling of the ab planes such as by the Josephson and/or proximity effects may mask the difference between $T_c^{ab}$ and $T_c^c$. Resistivity measurements as a function of T in the ab and c directions for a single crystal should be able to detect a difference in $T_c$.

**ACKNOWLEDGMENT**

I wish to express my heartfelt appreciation to Tom McMullen for bringing the Harshman and Mills paper to my attention, for helping me with the numerical calculations, for beneficial discussions, and most of all for his integrity, his valuable interest, his encouragement, and his support.

TABLE I. Wide Range of Experimental and Theoretical Transition Temperatures

| No. | Superconductor-fluid | $T_c^{exp}$ (K) | $T_c^{pred}$ (K) | $T_F^{2D}$ (K) | n ($10^{21}/cm^3$) | δ(Å) | $T_F^{3D}$ (K) |
|---|---|---|---|---|---|---|---|
| 1 | $^3He$ in $^4He$ | ? | $10^{-6}$ -$10^{-5}$ | - | - | - | - |
| 2 | H | ? | 30 x $10^{-6}$ | | $10^{-7}$ | | |
| 3 | $^3He$ | 0.0025 | 0.0028 | - | 23.6 | - | - |
| 4 | $^4He$ | 2.17 | 3 | - | 22 | - | - |
| 5 | $UPt_3$ | 0.53 | 0.68, 0.71, 0.74 | - | - | - | 124±5 |
| 6 | $[TMTSF]_2ClO_4$ | 1.2 | 2.31, 3.91, 6.35 | 72±24 | 0.38 ±0.17 | 13.275 | - |
| 7 | $TaS_2(Py)_{1/2}$ | 3.4 | 2.61, 3.60, 4.92 | 1710±360 | 12±3.6 | 12.02 | - |
| 8 | $PbMo_6S_8$ (Chevrel) | 12 | 43.4,56.1,68.7 | - | - | - | 1640±370 |
| 9 | κ-$[BEDT\text{-}TTF]_2Cu[NCS]_2$ | 10.5 | 7.26, 10.8, 15.3 | 213±57 | 0.31 ±0.09 | 15.24 | - |
| 10 | Tl | 2.39 | 30.8 | - | - | - | 9.46x$10^4$ |
| 11 | In | 3.4 | 32.6 | - | - | - | 10.0x$10^4$ |
| 12 | Sn | 3.7 | 38.4 | - | - | - | 11.8x$10^4$ |
| 13 | Hg | 4.15 | 27.0 | - | - | - | 8.29x$10^4$ |
| 14 | Pb | 7.19 | 35.8 | - | - | - | $11.0x10^4$ |
| 15 | Nb | 9.2 | 20.1 | - | - | - | $6.18x10^4$ |
| 16 | $Nb_3Sn$ | 17.9 | 2.44, 2.64, 2.84 | - | - | - | 8100±620 |
| 17 | $BaPb_{0.75}Bi_{0.25}O_3$ | 11 | 11.4, 12.4, 13.4 | - | - | - | 242±20 |
| 18 | $Ba_{0.6}K_{0.4}BiO_3$ | 32 | 55.9, 62.1, 68.3 | - | - | - | 1210±120 |
| 19 | $La_{1.9}Sr_{0.1}CuO_4$ | 33 | 45.8, 50.8, 55.8 | 510±50 | ~0.5 | 6.61 | - |
| 20 | $La_{1.875}Sr_{0.125}CuO_4$ | 36 | 33.4, 56.2, 61.7 | 710±7 | ~1.0 | 6.61 | - |
| 21 | $La_{1.85}Sr_{0.15}CuO_4$ | 39 | 43.0, 49.7, 57.3 | 1090±100 | 5.2 ±0.8 | 6.62 | - |
| 22 | $YBa_2Cu_3O_{6.67}$ | 60 | 49.8, 61.3, 79.4 | 710±70 | 1.1 ±0.4 | 5.87 | - |
| 23 | $YBa_2Cu_4O_8$ | 80 | 64.3, 74.1, 87.8 | 1360±140 | 2.8 ±0.44 | 6.81 | - |
| 24 | $YBa_2Cu_3O_7$ | 92 | 71.8, 79.9, 89.9 | 2290±100 | 16.9 ±3.4 | 5.84 | - |
| 25 | $HoBa_2Cu_4O_8$ | 80 | 131, 145, 159 | 2070±200 | ~1.3 | 6.82 | - |
| 26 | $Bi_2Sr_2CaCu_2O_8$ | 89 | 37.6, 43.2, 77.1 | 970±390 | 3.5 ±1.8 | 7.73 | - |
| 27 | $(Bi_{1.6}Pb_{0.4})Sr_2Ca_2Cu_3O_{10}$ | 107 | 29.5, 42.8, 61.1 | 1140±270 | 3.4 ±1.2 | 9.27 | - |
| 28 | $(Tl_{0.5}Pb_{0.5})Sr_2CaCu_2O_7$ | 80 | 75.5, 89.6, 103 | 1440±140 | 2.8 ±0.5 | 6.05 | - |
| 29 | $Tl_2Ba_2CaCu_2O_8$ | 99 | 41.0, 49.6, 61.2 | 1180±110 | 4.9 ±1.5 | 7.33 | - |
| 30 | $Tl_2Ca_2Ba_2Cu_3O_{10}$ | 125 | 53.5, 66.1, 84.4 | 1830±190 | 4.2 ±1.5 | 8.97 | - |

Where three numbers are shown for $T_c^{pred}$, these are the minimum, the mean, and the maximum predicted transition temperatures obtainable from the data. The mean value does not always lie symmetrically between the minimum and maximum values.